\documentclass[a4paper,12pt]{article}
\usepackage{jheppub}
\usepackage[dvipsnames]{xcolor}
\usepackage{pgfplots}
\usepackage[T1]{fontenc}
\usepackage[]{slashed}
\usepackage[]{bm}     
\usepackage{physics}
\usepackage{dsfont}

\usepackage[compat=1.1.0]{tikz-feynman}
\usepackage{tikz}
\usetikzlibrary{calc,decorations.markings}

\usepackage{diagrams}

\usepackage[force]{feynmp-auto}

\usepackage{hyperref}
\hypersetup{colorlinks=true,linkcolor=magenta,anchorcolor=green,citecolor=cyan,filecolor=black,menucolor=black,urlcolor=brown}
\makeatletter

\usepackage{lipsum}

\DeclareMathOperator\arccosh{arccosh}

\DeclareMathOperator\Li{Li}

\usepackage{graphicx}
\usepackage{epstopdf}
\usepackage{bbm,amsmath,graphicx,amssymb,amsfonts,amsthm}

\def\sq[#1,#2]{\left[#1\,#2\right]}
\def\an[#1,#2]{\left\langle#1\,#2\right\rangle}

\def\an[#1,#2]{\left\langle#1\,#2\right\rangle}
\def\aq[#1,#2,#3]{\left\langle#1|#2|#3\right]}
\def\qa[#1,#2,#3]{\left[#1|#2|#3\right\rangle}
\def\sq[#1,#2]{\left[#1\,#2\right]}
\def\spa#1.#2{\left\langle#1\,#2\right\rangle}
\def\spab[#1,#2,#3]{\left\langle#1|#2|#3\right]}
\def\spba[#1,#2,#3]{\left[#1|#2|#3\right\rangle}
\def\spb#1.#2{\left[#1\,#2\right]}

\def\Ttrma(#1,#2,#3,#4){{\rm tr}_{-}[\slash \!\!\!\;\!\! #1\slash  \!\!\!\;\!\! #2 \slash  \!\!\!\;\!\!#3\slash  \!\!\!\;\!\!#4]}
\def\Ttrmb(#1,#2,#3,#4,#5,#6){{\rm tr}_{-}[\slash \!\!\!\;\!\! #1\slash  \!\!\!\;\!\! #2 \slash  \!\!\!\;\!\!#3\slash  \!\!\!\;\!\!#4\slash  \!\!\!\;\!\!#5\slash  \!\!\!\;\!\!#6]}
\def\Ttrmc(#1,#2,#3,#4,#5,#6,#7,#8){{\rm tr}_{-}[\slash \!\!\!\;\!\! #1\slash  \!\!\!\;\!\! #2 \slash  \!\!\!\;\!\!#3\slash  \!\!\!\;\!\!#4\slash  
\!\!\!\;\!\!#5\slash  \!\!\!\;\!\!#6\slash  \!\!\!\;\!\!#7\slash  \!\!\!\;\!\!#8]}
\def\Dp(#1,#2){(#1\cdot #2)}

\def\triangleboxleft{\scalebox{.9}{$\triangleleft$}\kern-.1em\Box}
\def\triangleboxright{\Box\kern-.1em\scalebox{.9}{$\triangleright$}}
\def\dBox{\Box\kern-.1em\Box}
\def\dNPBoxs{\scalebox{.9}{$\bowtie$}\kern-.1em\Box}
\def\dNPBoxu{\Box\kern-.1em\scalebox{.9}{$\bowtie$}}

\def\beq{\begin{equation}}
\def\eeq{\end{equation}}
\def\bes{\begin{split}}
\def\ees{\end{split}}
\def\beqa{\begin{eqnarray}}
\def\eeqa{\end{eqnarray}}

\def\eeqa{\end{eqnarray}}
\def\ek[#1,#2]{(\varepsilon_{#1}\cdot k_{#2})}
\def\e[#1,#2]{(\varepsilon_{#1}\cdot \varepsilon_{#2})}
\def\s(#1,#2){{(\ell_#1\cdot\ell_#2)}}

\def\e{\epsilon}

\pgfdeclarelayer{bg}    % declare background layer
\pgfsetlayers{bg,main}  % set the order of the layers (main is the standard layer)

\definecolor{Mathematica}{HTML}{ed192d}

\usepackage{float}

\usetikzlibrary{shapes.misc}
\usetikzlibrary{arrows}
\usetikzlibrary{decorations.markings}
\usetikzlibrary{positioning,fit}
\usetikzlibrary{patterns}

\tikzset{cross/.style={cross out, draw=black, minimum size=2*(#1-\pgflinewidth), inner sep=0pt, outer sep=0pt},
	cross/.default={2pt}}

 \preprint{\vbox{\hbox{\hphantom{XXXX}IPhT-t21/037}\hbox{\hphantom{X}CERN-TH-2021-111}}}
      
\title{On an Exponential Representation \\of the Gravitational S-Matrix}

\author[a,b]{Poul H. Damgaard}
\author[a]{\!\!, Ludovic Plant\'e}
\author[c,d]{\!\!, Pierre Vanhove}

\affiliation[a]{Niels Bohr International Academy, Niels Bohr Institute, University of Copenhagen, Blegdamsvej 17, DK-2100 Copenhagen, Denmark}
\affiliation[b]{Theoretical Physics Department, CERN, 1211 Geneva 23, Switzerland}
\affiliation[c]{Universit\'e Paris-Saclay, CNRS, CEA, Institut de physique th\'eorique, 91191, Gif-sur-Yvette, France}
\affiliation[d]{National Research University Higher School of
  Economics, Russian Federation}


\abstract{An exponential representation of the S-matrix provides a natural framework for understanding the semi-classical limit of scattering amplitudes. 
While sharing some similarities with the eikonal formalism it differs from it in details. 
Computationally, rules are simple because pieces that must be subtracted are given by combinations of unitarity cuts.
Analyzing classical gravitational scattering to third Post-Minkowskian order in both maximal supergravity and Einstein
gravity we find agreement with other approaches, including the contributions from radiation reaction terms. The kinematical relation for
the two-body problem in isotropic coordinates follows immediately from this procedure, again with the inclusion of radiation reaction pieces up to third 
Post-Minkowskian order.}
\begin{document} 
\maketitle
\flushbottom
\newpage

\section{Introduction}

The surge of interest in using modern amplitude techniques to calculate the Post-Minkowskian
expansion of classical general relativity~\cite{Damour:2016gwp,Damour:2017zjx,Bjerrum-Bohr:2018xdl,Cheung:2018wkq,Kosower:2018adc,Bern:2019nnu,Antonelli:2019ytb,Cristofoli:2019neg,Bern:2019crd,Kalin:2019rwq,Bjerrum-Bohr:2019kec,Cristofoli:2020uzm,Parra-Martinez:2020dzs,DiVecchia:2020ymx,Damour:2020tta,DiVecchia:2021ndb,Bern:2021dqo,DiVecchia:2021bdo,Bjerrum-Bohr:2021vuf,Bjerrum-Bohr:2021din,Bini:2021gat,Bautista:2021wfy,Cristofoli:2021vyo,Herrmann:2021lqe,Herrmann:2021tct,Mougiakakos:2021ckm,Jakobsen:2021smu}
has led to a renewed focus on the eikonal formalism in relativistic quantum field theory. This eikonal approach is based on an exponentiation of the scattering amplitude
in impact-parameter space and it is particularly well suited for understanding classical gravity from a field theoretic setting~\cite{'tHooft,Amati:1987wq,Amati:1987uf,Amati:1990xe,Amati:1992zb,Amati:1993tb,Kabat:1992tb,Akhoury:2013yua,Luna,Collado:2018isu,Paolodue,DAppollonio:2010krb,DAppollonio:2013mgj,Paolotre,Paoloquatro,Bern:2020gjj}. Nevertheless, as has been known for long (see, $e.g.$, Appendix B of ref.~\cite{Amati:1990xe} as well as ref.~\cite{Ciafaloni:2014esa}), 
the precise definition of the eikonal phase becomes subtle at higher orders. What in a first approximation can be treated as insignificant
contributions from small transverse momenta $q^2$ become important and have to be taken into account. Such terms may appear as being of quantum origin but since they 
mix with terms from the Laurent expansion in $\hbar$ of higher orders in the perturbative expansion
they cannot be ignored. Although there is little doubt that the eikonal formalism can be pushed to arbitrarily high order by a careful analysis 
(the small-angle classical scattering regime exists and should be computable systematically order by order in the coupling), it nevertheless suggests that it may be useful to 
pursue alternative strategies as well.

A different approach to the semi-classical limit of the $S$-matrix is the WKB formalism.  While well-known in the context of non-relativistic quantum mechanics, where it provides a systematic semi-classical expansion of the wave function through its connection
to the classical Hamilton-Jacobi formalism, little attention has so far been paid to the WKB framework in relativistic quantum field theory. 
Motivated by the recent proposal of Bern et al. ~\cite{Bern:2021dqo} we here wish to explore the possibility
of using the WKB limit of relativistic quantum field theories to derive in an alternative manner the classical scattering of two massive objects in general relativity from
scattering amplitudes. 

The eikonal formalism has issues of complexity at several layers. First, while scattering amplitudes are conveniently computed in the plane-wave basis of momentum
space, the eikonal lives in impact-parameter space. To obtain the needed eikonal exponentiation, one must carefully separate, order by order, those terms that go into the
exponent and those terms that remain as prefactors at the linear level. Second, after exponentiation in impact-parameter space one must apply the inverse transformation
and seek from it two crucial ingredients: (1) the correct identification of the transverse momentum transfer $\vec{q}$ in the center of mass frame and (2) the correct identification
of the scattering angle from the saddle point. At low orders in the eikonal expansion this procedure works well but it hinges on the  impact-parameter
transformation being able to undo the convolution product of the momentum-space representation. When $q^2$-corrections are taken into account it is well known
that this procedure requires amendments. This motivates why an alternative pathway, such as the one proposed in ref.~\cite{Bern:2021dqo} which we find is rooted in the 
WKB approximation rather than the eikonal {\em per se} should be investigated.

We begin our paper by first introducing the exponential representation of the $S$-matrix and we then proceed to develop a method that differs from that of ref.~\cite{Bern:2021dqo}. 
We shall check our proposal to third Post-Minkowskian order for both maximal supergravity and Einstein gravity, and we shall also point out a simple link
to the potential from the Hamiltonian formalism. This has all classical contributions to the order we work, including radiation reaction parts.

\section{Exponential representation of the $S$-matrix}

Conventionally, the $S$-matrix is expanded in the form
\begin{equation}
  \hat{S}=1 + \frac{i}{\hbar}\hat{T}  
  \label{Born}
\end{equation}
where $\hat{T}$ is the scattering operator. Restricting ourselves at first to two-particle scalar scattering with incoming and outgoing momenta $p_1, p_2$ and $p_1',p_2'$, respectively,
matrix elements of $\hat{T}$ define for us the scattering amplitude $M(p_1,p_2,p_1',p_2') $ through
\begin{equation}
  \langle p_1',p_2'| \hat{T} | p_1, p_2 \rangle  ~=~  (2\pi \hbar)^D\delta^{(D)}(p_1+p_2-p_1'-p_2') M(p_1,p_2,p_1',p_2')  ~.
\end{equation}
At this stage it is convenient to introduce also the exchanged four-momentum $q$ through
$p_1'=p_1+q$ and $p_2'=p_2-q$. The  center-of-mass kinematics  reads
\begin{equation}
  p_1=(E_1(p),\vec p), \qquad p_2=(E_2(p),-\vec p)
\end{equation}
so that $q=p_1-p_1'=-p_2+p_2'$ with $p_1^2=(p_1')^2=m_1^2$ and $p_2^2=(p_2')^2=m_2^2$, where we have introduced the two masses $m_1$ and $m_2$.

Because the $S$-matrix is unitary, $\hat{S}^{\dagger}\hat{S} = 1$, a more natural parametrization might appear to be
\beq
\hat{S} ~=~ \exp\left[{i\over\hbar}\hat{N}\right]
\label{Ndef}
\eeq
where, in contrast to $\hat{T}$, the operator $\hat{N}$ is Hermitian:
$\hat{N} = \hat{N}^{\dagger}$. The  eigenvalues of  $S$ are then manifestly
pure phases, defined mod $2\pi$.

Standard quantum field theory is set up to evaluate matrix elements of the $\hat{T}$-matrix, providing us with the conventional Born series of
perturbation theory. However, it is clearly straightforward to relate $\hat{N}$ to $\hat{T}$ at the operator level. Everything can be formulated in general terms but since
we are interested in the application to classical gravity we will be specific here. In perturbation theory of gravity we assume that both $\hat{T}$ and $\hat{N}$ can be 
expanded in Newton's constant $G_N$. When an odd number of gravitons are emitted the expansion of the fundamental $\hat{T}$-matrix will include powers
of $\sqrt{G_N}$ and for consistency this must also be the power series expansion of $\hat{N}$. Without having to write out explicitly the precise expansion of
the $S$-matrix in terms of the Dyson expansion, we can therefore in a compact notation write\footnote{We do not include the lowest-order radiative
terms of order $G_N^{1/2}$ because they are not needed for our discussion of classical two-body scattering to the order we consider below. Radiative terms contribute also 
to even powers in $G_N$ but we choose here to absorb them into the non-radiative terms without explicit labelling.}
\begin{eqnarray}
\hat{T}  &=&  G_N\hat{T}_0 + G_N^{3/2}\hat{T}_0^{\rm rad} + G_N^2\hat{T}_1 + G_N^{5/2}\hat{T}_1^{\rm rad} + G_N^3\hat{T}_2 + ... \cr
\hat{N}  &=&  G_N\hat{N}_0 + G_N^{3/2}\hat{N}_0^{\rm rad} + G_N^2\hat{N}_1 + G_N^{5/2}\hat{N}_1^{\rm rad} + G_N^3\hat{N}_2 + ... 
\end{eqnarray}
which leads to the operator identifications
\begin{eqnarray}\label{e:TtoN}
\hat{T}_0  &=&  \hat{N}_0 \cr
\hat{T}_0^{\rm rad} &=& \hat{N}_0^{\rm rad} \cr
\hat{T}_1  &=&  \hat{N}_1 + \frac{i}{2\hbar}\hat{N}_0^2 \cr
\hat{T}_1^{\rm rad} &=& \hat{N}_1^{\rm rad} +\frac{i}{2\hbar}(\hat{N}_0\hat{N}_0^{\rm rad} + \hat{N}_0^{\rm rad}\hat{N}_0) \cr
\hat{T}_2  &=&  \hat{N}_2  +\frac{i}{2\hbar}(\hat{N}_0^{\rm rad})^2 + \frac{i}{2\hbar}(\hat{N}_0\hat{N}_1 + \hat{N}_1\hat{N}_0)- \frac{1}{3! \hbar^2}\hat{N}_0^3
\end{eqnarray}
and so on for higher orders. We have here kept explicit factors of $\hbar$. The ordering in terms of $\hbar$ clearly becomes meaningful only once we evaluate matrix 
elements: loop contributions to $\hat{T}$ matrix elements will produce a Laurent series in $\hbar$, thus reshuffling the apparent counting at the operator level. 

Iteratively, we can relate $\hat{N}_i$ to $\hat{T}_i$ and lower-order $\hat{T}$'s. In detail,
\begin{eqnarray}\label{e:NtoT}
\hat{N}_0  &=&  \hat{T}_0 \cr
\hat{N}_0^{\rm rad} &=& \hat{T}_0^{\rm rad} \cr
\hat{N}_1  &=&  \hat{T}_1 - \frac{i}{2\hbar}\hat{T}_0^2 \cr
\hat{N}_1^{\rm rad} &=& \hat{T}_1^{\rm rad} -\frac{i}{2\hbar}(\hat{T}_0\hat{T}_0^{\rm rad} + \hat{T}_0^{\rm rad}\hat{T}_0) \cr
\hat{N}_2  &=&  \hat{T}_2  - \frac{i}{2\hbar}(\hat{T}_0^{\rm rad})^2 - \frac{i}{2\hbar}(\hat{T}_0\hat{T}_1 + \hat{T}_1\hat{T}_0) - \frac{1}{3 \hbar^2}\hat{T}_0^3
\end{eqnarray}
and similarly for higher orders.

We now proceed to consider matrix elements of the $S$-matrix in momentum space. In addition, and in order to elucidate the structure, it is convenient to also
insert completeness relations whenever operator multiplications occur. We write the completeness relation symbolically as
\beq
\sum_n | n \rangle\langle n |  ~=~ \mathbb I
\eeq
where the sum over $n$ runs over the complete set of accessible states. Specifically, and again with a view towards the scattering of two heavy objects in gravity, 
the sum takes the explicit form
\begin{multline}\label{e:complet}
\mathbb I ~=~ \sum_{n=0}^{\infty} \frac{1}{n!} \int \frac{d^{(D-1)}k_1}{(2\pi\hbar)^{(D-1)}} \frac{1}{2E_{k_{1}}}  \frac{d^{(D-1)}k_2}{(2\pi\hbar)^{(D-1)}} \frac{1}{2E_{k_{2}}} 
\frac{d^{(D-1)}\ell_1}{(2\pi\hbar)^{(D-1)}} \frac{1}{2E_{\ell_{1}}}  \cdots \frac{d^{(D-1)}\ell_n}{(2\pi\hbar)^{(D-1)}} \frac{1}{2E_{\ell_{n}}} \cr\times
| k_1, k_2; \ell_1, \ldots \ell_n \rangle\langle k_1, k_2; \ell_1, \ldots \ell_n |,
\end{multline}
where the state $n=0$ corresponds to just the two massive scalars, the $n=1$ state to one graviton in addition, etc.  As is well known, the sum over states is the Lorentz invariant
phase space. This is important because the completeness relation will hence relate expressions to cuts and, therefore, unitarity. When we use the complete set of
states to saturate the matrix elements to any order in $G$ it becomes immediately obvious which terms include radiative parts.

Considering just two-body scattering, matrix elements of $\hat{N}$ in momentum space now read:
\begin{eqnarray}
\langle p_1',p_2' | \hat{N}_0  | p_1, p_2 \rangle &=&  \langle p_1',p_2' | \hat{T}_0 | p_1, p_2 \rangle, \cr
\langle p_1',p_2' | \hat{N}_0^{\rm rad} | p_1, p_2 \rangle &=& 0, \cr
\langle p_1',p_2' | \hat{N}_1  | p_1, p_2 \rangle &=&  \langle p_1',p_2' | \hat{T}_1 | p_1, p_2 \rangle - \frac{i}{2\hbar} \sum_n  \langle p_1',p_2' |  \hat{T}_0 | n \rangle\langle n | \hat{T}_0 | p_1, p_2 \rangle,  \cr
\langle p_1',p_2' | \hat{N}_1^{\rm rad}  | p_1, p_2 \rangle &=& 0, \cr
\langle p_1',p_2' | \hat{N}_2  | p_1, p_2 \rangle &=&  \langle p_1',p_2' | \hat{T}_2 | p_1, p_2 \rangle  - 
\frac{i}{2\hbar} \sum_n  \langle p_1',p_2' |  \hat{T}_0^{\rm rad} | n \rangle\langle n | \hat{T}_0^{\rm rad} | p_1, p_2 \rangle \cr
&& - \frac{i}{2\hbar}\sum_n(\langle p_1',p_2' | \hat{T}_0| n \rangle\langle n | \hat{T}_1 | p_1, p_2 \rangle+ \langle p_1',p_2' | \hat{T}_1| n \rangle\langle n | \hat{T}_0 | p_1, p_2 \rangle)
\cr && - \sum_{n,m}\frac{1}{3 \hbar^2}\langle p_1',p_2' | \hat{T}_0 | n \rangle\langle n | \hat{T}_0 | m \rangle\langle m | \hat{T}_0 | p_1, p_2 \rangle.
\label{Nmatrixelements}
\end{eqnarray}

We straightforwardly infer some basic facts from these expressions. First, the matrix element of $\hat{T}_0$, the leading term of the $\hat{T}$-matrix, must be real. Second,
because also the matrix element of $\hat{N}_1$ is real we conclude that
\beq
\Im[\langle p_1',p_2' | \hat{T}_1 | p_1, p_2 \rangle] ~=~   \frac{1}{2\hbar} \sum_n  \langle p_1',p_2' |  \hat{T}_0 | n \rangle\langle n | \hat{T}_0 | p_1, p_2 \rangle 
\label{firstUnitarity}
\eeq
where we have abbreviated the completeness relation~\eqref{e:complet}. 
This relation is the leading-order expression of unitarity,
\beq
\hat{T} - \hat{T}^\dagger ~=~ \frac{i}{\hbar}\hat{T}\hat{T}^{\dagger} ~.
\label{unitarityT}
\eeq
On account of coupling constant counting, the right hand side of eq.~(\ref{firstUnitarity}) is saturated by elastic unitarity so that
\begin{multline}\label{firstUnitarity2}
  \Im[\langle p_1',p_2' | \hat{T}_1 | p_1, p_2 \rangle] \cr
=  \frac{1}{2\hbar} \int \frac{d^{(D-1)}k_1}{(2\pi\hbar)^{(D-1)}} \frac{1}{2E_{k_{1}}}   \frac{d^{(D-1)}k_2}{(2\pi\hbar)^{(D-1)}} \frac{1}{2E_{k_{2}}} 
\langle p_1',p_2' |  \hat{T}_0 | k_1,k_2 \rangle\langle k_1,k_2 | \hat{T}_0 | p_1, p_2 \rangle 
\end{multline}
is exact.

The next step where unitarity allows us to understand the separation into real and imaginary parts is in the evaluation of the matrix element of $\hat{N}_2$. Although we are 
here restricting ourselves to elastic scattering (it can easily be extended to include radiation following the general steps above) radiation appears indirectly because of
unitarity. From eq.~(\ref{e:NtoT}) it follows that the matrix element of $\hat{T}_0^{\rm rad}$, the leading term of the $\hat{T}$-matrix that includes radiation, must be real.
Let us now consider the unitarity relation~(\ref{unitarityT}) at order $G_N^3$:
\beq
\hat{T}_2 - \hat{T}_2^{\dagger}  ~=~ \frac{i}{\hbar} \left[(\hat{T}_0^{\rm rad})^2 + \hat{T}_0\hat{T}_1^{\dagger} + \hat{T}_1\hat{T}_0\right] ~.
\eeq
What is inside the parenthesis on the right hand side of this equation is indeed Hermitian. 
Inserting complete sets of states, we see that the first term is saturated by an intermediate state of two scalars
and one graviton while the remaining terms are saturated by just the
two-particle state of the scalars, thus
\begin{multline}
2\hbar  \Im[\langle 2| \hat{T}_2 | 2 \rangle] \cr
  = \int\frac{d^{(D-1)}k_1}{(2\pi\hbar)^{(D-1)}} \frac{1}{2E_{k_{1}}}  \frac{d^{(D-1)}k_2}{(2\pi\hbar)^{(D-1)}} \frac{1}{2E_{k_{2}}} 
\frac{d^{(D-1)}\ell}{(2\pi\hbar)^{(D-1)}2E_\ell}\langle
2| \hat{T}_0^{\rm rad}|k_1,k_2;\ell\rangle\langle k_1,k_2;\ell | \hat{T}_0^{\rm
  rad}|2\rangle \cr
 +  \int  \frac{d^{(D-1)}k_1}{(2\pi\hbar)^{(D-1)}}
\frac{1}{2E_{k_{1}}}  \frac{d^{(D-1)}k_2}{(2\pi\hbar)^{(D-1)}}
\frac{1}{2E_{k_{2}}}
\Big(
\langle 2 | \hat{T}_0| k_1,k_2\rangle\langle
k_1,k_2|\hat{T}_1^{\dagger}|2\rangle \cr+ \langle 2| \hat{T}_1|
k_1,k_2\rangle\langle k_1,k_2| \hat{T}_0 | 2\rangle\Big).
\label{secondUnitarity}
\end{multline}
Also this equation is exact to all orders in $\hbar$. Comparing with eq.~(\ref{Nmatrixelements}) we see how the imaginary part from $\hat{T}_2$ is cancelled partly
by the radiation term, partly by the iteration of the real part of $\hat{T}_1$. Of course, $\hat{T}_1$ also has an imaginary part which indeed contributes, through
the shown iteration, to $\hat{N}_2$. Also in the relation for $\hat{N}_2$ in eq.~(\ref{Nmatrixelements}) the sums over states are saturated by two scalars and one
graviton for the term involving $\hat{T}_0^{\rm rad}$ and just two scalars in the remaining terms.

At higher orders these relations get increasingly complicated although
of course the two body matrix elements of $\hat{N}$ are always
real.\footnote{ The reality of the two-particle elements is a
  consequence of the hermiticity of the operator $(\langle p_1,p_2| \hat
  N|p_3,p_4\rangle)^*=\langle p_3,p_4|\hat N^\dagger|p_1,p_2\rangle=
  \langle p_3,p_4|\hat N|p_1,p_2\rangle$  and the time
  reversibility of the amplitudes $(\langle p_1,p_2| \hat
  N|p_3,p_4\rangle)^*=
  \langle p_1,p_2|\hat N|p_3,p_4\rangle$. }
  reality of the matrix elements  For each $\hat{N}_i$ the
subtractions of eq.~(\ref{e:NtoT}) remove precisely all imaginary parts from the matrix element of $\hat{T}_i$. 

The expansions~(\ref{Nmatrixelements}) have nice diagrammatic interpretations in terms of unitarity cuts. At one-loop order,
\begin{equation}
N_1=M_1- {i\over2}\qquad\begin{tikzpicture}[baseline=(z)]
	\begin{feynman}[inline=(z)]
	\tikzfeynmanset{every vertex={empty dot,minimum size=0mm}}
        \vertex(z);
%	\tikzfeynmanset{every vertex=dot}
	\vertex [above left=.4cm and 1.6cm of z] (x1);
	\vertex [below left=.4cm and 1.6cm of z] (x2);
	\vertex [above right=.4cm and 1.6cm of z] (x4);
	\vertex [below right=.4cm and 1.6cm of z] (x3);
	\vertex [left=1cm of z] (z1);
	\vertex [right=1cm of z] (z2);
	\vertex [above=1.1cm of z] (z3);
          \vertex [below=1.1cm of z] (z4);
	\diagram* {
                (x1) -- (x4);
                (x2) -- (x3);
                (z3) --[scalar,color=red]  (z4);
	};
	\end{feynman} 
        \draw[gray,fill=gray] (z1) circle (.53cm)  node[black,below=.5cm]{$M_0$};
\draw[gray,fill=gray] (z2) circle (.53cm)  node[black,below=.5cm]{$M_0$};;
	\end{tikzpicture}
\end{equation}
and at two-loop order,
\begin{align}
N_2&=M_2- {i\over2}\qquad\begin{tikzpicture}[baseline=(z)]
	\begin{feynman}[inline=(z)]
	\tikzfeynmanset{every vertex={empty dot,minimum size=0mm}}
        \vertex(z);
%	\tikzfeynmanset{every vertex=dot}
	\vertex [above left=.4cm and 1.6cm of z] (x1);
	\vertex [below left=.4cm and 1.6cm of z] (x2);
	\vertex [above right=.4cm and 1.6cm of z] (x4);
	\vertex [below right=.4cm and 1.6cm of z] (x3);
	\vertex [ left=.4cm of z] (y5);
	\vertex [ right=.4cm of z] (y6);
	\vertex [left=1cm of z] (z1);
	\vertex [right=1cm of z] (z2);
	\vertex [above=1.1cm of z] (z3);
          \vertex [below=1.1cm of z] (z4);
	\diagram* {
                (x1) -- (x4);
                (x2) -- (x3);
                (y5) -- [gluon] (y6);
                (z3) --[scalar,color=red]  (z4);
	};
	\end{feynman} 
        \draw[gray,fill=gray] (z1) circle (.53cm)  node[black,below=.5cm]{$M_0^{\rm rad}$};
\draw[gray,fill=gray] (z2) circle (.53cm)  node[black,below=.5cm]{$M_0^{\rm rad}$};
	\end{tikzpicture}\\
\nonumber &-{i\over2}\left(\begin{tikzpicture}[baseline=(z)]
	\begin{feynman}[inline=(z)]
	\tikzfeynmanset{every vertex={empty dot,minimum size=0mm}}
        \vertex(z);
%	\tikzfeynmanset{every vertex=dot}
	\vertex [above left=.4cm and 1.6cm of z] (x1);
	\vertex [below left=.4cm and 1.6cm of z] (x2);
	\vertex [above right=.4cm and 1.6cm of z] (x4);
	\vertex [below right=.4cm and 1.6cm of z] (x3);
	\vertex [left=1cm of z] (z1);
	\vertex [right=1cm of z] (z2);
	\vertex [above=1.1cm of z] (z3);
          \vertex [below=1.1cm of z] (z4);
	\diagram* {
                (x1) -- (x4);
                (x2) -- (x3);
                          (z3) --[scalar,color=red]  (z4);
	};
	\end{feynman} 
        \draw[gray,fill=gray] (z1) circle (.53cm)  node[black,below=.5cm]{$M_0$};
\draw[gray,fill=gray] (z2) circle (.53cm)  node[black,below=.5cm]{$M_1$};
	\end{tikzpicture}+
\begin{tikzpicture}[baseline=(z)]
	\begin{feynman}[inline=(z)]
	\tikzfeynmanset{every vertex={empty dot,minimum size=0mm}}
        \vertex(z);
%	\tikzfeynmanset{every vertex=dot}
	\vertex [above left=.4cm and 1.6cm of z] (x1);
	\vertex [below left=.4cm and 1.6cm of z] (x2);
	\vertex [above right=.4cm and 1.6cm of z] (x4);
	\vertex [below right=.4cm and 1.6cm of z] (x3);
	\vertex [left=1cm of z] (z1);
	\vertex [right=1cm of z] (z2);
	\vertex [above=1.1cm of z] (z3);
          \vertex [below=1.1cm of z] (z4);
	\diagram* {
                (x1) -- (x4);
                (x2) -- (x3);
                (z3) --[scalar,color=red]  (z4);
	};
	\end{feynman} 
        \draw[gray,fill=gray] (z1) circle (.53cm)  node[black,below=.5cm]{$M_1$};
\draw[gray,fill=gray] (z2) circle (.53cm)  node[black,below=.5cm]{$M_0$};
	\end{tikzpicture}\right)
-{1\over3}\ \begin{tikzpicture}[baseline=(z)]
	\begin{feynman}[inline=(z)]
	\tikzfeynmanset{every vertex={empty dot,minimum size=0mm}}
        \vertex(z);
%	\tikzfeynmanset{every vertex=dot}
	\vertex [above left=.4cm and 1.6cm of z] (x1);
	\vertex [below left=.4cm and 1.6cm of z] (x2);
	\vertex [above right=.4cm and 3.5cm of z] (x4);
	\vertex [below right=.4cm and 3.5cm of z] (x3);
	\vertex [left=1cm of z] (z1);
	\vertex [right=1cm of z] (z2);
	\vertex [right=3cm of z] (z3);
	\vertex [above=1.1cm of z] (z4);
          \vertex [below=1.1cm of z] (z5);
	\vertex [above right=1.1cm and 2cm of z] (z6);
          \vertex [below right=1.1cm and 2cm of z] (z7);
	\diagram* {
                (x1) -- (x4);
                (x2) -- (x3);
                          (z4) --[scalar,color=red]  (z5);
   (z6) --[scalar,color=red]  (z7);
	};
	\end{feynman} 
        \draw[gray,fill=gray] (z1) circle (.53cm)  node[black,below=.5cm]{$M_0$};
\draw[gray,fill=gray] (z2) circle (.53cm)  node[black,below=.5cm]{$M_0$};
\draw[gray,fill=gray] (z3) circle (.53cm)  node[black,below=.5cm]{$M_0$};
	\end{tikzpicture}.
\end{align}
The cut involving $\hat{T}_0^{\rm rad}$ was first computed in ref.~\cite{DiVecchia:2021bdo}. We see here how this imaginary part of the matrix element
of $\hat{T}_2$ is automatically subtracted off as dictated by unitarity. Similar radiative cancellations occur at higher orders. It is interesting to compare with the
manner in which terms exponentiate in the eikonal approach where the additional subtractions of imaginary parts of eq.~(\ref{secondUnitarity}) occur as part
of the lower-point iterations, instead, as here, in one go because of unitarity.

To summarize this part, the matrix element of any $\hat{N}_i$ is manifestly real. 
In addition, all iterations of lower-point amplitudes and their corresponding super-classical terms are automatically
subtracted off so that matrix elements of $\hat{N}$ admit a semi-classical limit where $\hbar \to 0$. Such an object could be expected to have a classical meaning
from analytical mechanics. Resembling the method of Born subtractions that lead to the classical potential from the solution of the Lippmann-Schwinger equation
\cite{Cristofoli:2019neg} one might guess that it could define in an alternative manner the classical potential. Although Born subtractions share the properties of removing 
all imaginary parts from the amplitude and of ensuring the existence of a semi-classical limit $\hbar \to 0$, such an identification does not hold beyond leading orders.

We can instead get a hint from the identification of the exponential representation of the two-body $S$-matrix in momentum space with the semi-classical WKB approximation.  
As shown in ref.~\cite{Eu}, the Schr\"odinger-like equation satisfied by the $U$-matrix (the limit of which provides the $S$-matrix) leads to 
a systematic semi-classical expansion of the momentum-space $S$-matrix that is a direct analog of the semi-classical WKB expansion of the wave function in 
non-relativistic quantum mechanics. The pertinent Hamilton-Jacobi equation of the leading-order approximation is in a canonically transformed form compared to
the more commonly used in analytical mechanics where, up to a sign, coordinates and momenta are swapped. In quantum field theory this is
implemented by a Fourier transform in the usual way. With this identification, the two-body $S$-matrix in momentum space admits a classical limit in terms of
the WKB phase shift, 
\beq
\langle p_1',p_2' | \hat{S}| p_1, p_2 \rangle ~\sim~ e^{2i\delta/\hbar}
\eeq
where 
\beq
\delta =  J \frac{\pi}{2} + \int_{r_{m}}^{\infty} dr(p_r - p_{\infty}) - p_{\infty}r_m
\eeq
is the radial action. The first term corresponds to the free motion, $p_r$ is radial momentum, $r_m$ is the classical turning point (minimal distance in the case of scattering), 
and $p_{\infty}$ is the 3-momentum of either particle in the center of
mass frame, evaluated at infinity\footnote{This form of the radial
  action appears to depend on the classical turning point $r_m$. In
  fact, because this turning point is uniquely fixed by the angular
  momentum $J$ and initial momentum $p_\infty$ the final result does
  not depend on it. Starting with the main
equation for the scattering angle $\chi$~\cite{Bjerrum-Bohr:2019kec},
\begin{equation}
\chi=\sum_{k=1}^{\infty} \frac{2 b}{k!}\int_0^\infty du
\Big(\frac{d}{du^2} \Big)^k \left[\frac{V_{\rm eff}^k(\sqrt{u^2+b^2}) (u^2+b^2)^{k-1}}{p_{\infty}^{2k}}\right]
\end{equation}
we find that the radial action is
\begin{equation}
\int_{r_m}^{+\infty} dr p_r= -\frac{\pi J}{2}-\frac{1}{2}\sum_{k=1}^{\infty} \frac{1}{k!}  \int_0^\infty du \Big(\frac{d}{du^2} \Big)^{k-1} \left[\frac{V_{\rm eff}^k(\sqrt{u^2+b^2}) (u^2+b^2)^{k-1}}{p_{\infty}^{2k-1}}\right]
\end{equation}
up to a constant which is irrelevant here. The first term is just the free radial action.}. This simple expression is valid in the elastic channel.

The conventional approach to linking the scattering matrix $\hat{T}$ with the classical trajectory of this two-body problem would be to transform to impact-parameter
space as in the eikonal approximation. This unwinds the already beautifully exponentiated form of the $S$-matrix as in eq.~(\ref{Ndef}) and would essentially take us back to
the complications of the eikonal expansion beyond the first leading orders. Instead, one could ask if it might be possible to start the analysis directly with the operator
$\hat{N}$, essentially viewing it as the operator of the phase shift itself, $i.e.$ here the radial action. Instead of the $S$-matrix itself, we thus consider matrix elements
of the $\hat{N}$-operator. This seems to be what lies behind the interesting proposal of a recent paper 
by  Bern et al. in ref.~\cite{Bern:2021dqo}, although it is phrased there with somewhat different terminology. An immediate issue is how to go from momentum space to 
impact-parameter space. In the above paper it is proposed to do this by means of a $(D-2)$-dimensional Fourier transform, as in the eikonal formalism, in conjunction with a subtraction formalism with roots in effective field theory.  This has been employed to compute up to fourth Post-Minkowskian order in what is technically known as the potential region~\cite{Bern:2021dqo}. We shall here view the same idea from the slightly different perspective outlined above and with what we believe is a simpler 
and more general formalism. We will check the results up to third Post-Minkowskian order and will simultaneously show that this method is not limited to the potential region.

\section{Unitarity cuts and velocity cuts}

In practical terms, the expansions~(\ref{Nmatrixelements}) are useful for the same reason the effective field theory subtractions do their job in ref.~\cite{Bern:2021dqo}:
they tell us which parts of the matrix elements of $\hat{T}_i$ we do not have to compute since we know that they will be subtracted off anyway.

Because the idea of computing the matrix element of the exponentiated operator $\hat{N}$ is new, we nevertheless find it illuminating to work through the
computations and thus show how the subtractions of cut diagrams leave us precisely with the desired object up to third order in the 
Post-Minkowskian expansion. We will use the same method and the same basis of integrals as in refs.~\cite{Bjerrum-Bohr:2021vuf,Bjerrum-Bohr:2021din} to which
we refer for further details.

As a warm-up exercise, we first go through the one-loop calculation in all details. Already at this order we will need to carefully keep terms that appear as
being of quantum origin but which, nevertheless, will contribute to classical physics at higher orders. In fact, we shall provide an all-order result in $q^2$
which will be needed at all subsequent orders in the Post-Minkowskian expansion.

As our starting point, consider the box integral in $D=4-2\epsilon$ dimensions
\begin{equation}
I_{\Box}^{s}  =  -\frac{1}{4}\int \frac{d^D k}{(2\pi \hbar)^D} \frac{\hbar^5}{(p_1 \cdot k+i \epsilon)(p_2 \cdot k-i \epsilon)k^2 (k+q)^2}.
\end{equation}
In this expression we neglect the $k^2$ terms of the massive
propagators as they vanish in what corresponds to a soft expansion of
$I_{\Box,s}$ in powers of $q$. This is explained in Appendix~\ref{sec:real}.
As in~\cite{Bjerrum-Bohr:2021vuf,Bjerrum-Bohr:2021din}, we write $q$
as $q= \hbar |\vec{\underline q} |  u_q$ where $u_q$ is a unit space-like vector ($u_q^2=-1$). 
Doing the change of variable $k \rightarrow \hbar|\vec{\underline q}| k$ we get
\begin{equation}
I_{\Box}^{s}=-\frac{|\vec{\underline q}|^{D-6}}{4 \hbar}\int \frac{d^D k}{(2\pi)^D} \frac{1}{(p_1 \cdot k+i \varepsilon)(p_2 \cdot k-i \varepsilon)k^2 (k+u_q)^2}.
\end{equation}
The integral can be written equivalently in terms of $p_3$ and $p_4$ as 
\begin{equation}
I_{\Box}^{s}=-\frac{|\vec{\underline q}|^{D-6}}{4\hbar}\int \frac{d^D k}{(2\pi)^D} \frac{1}{(p_3 \cdot k-i \varepsilon)(p_4 \cdot k+i \varepsilon)k^2 (k+u_q)^2}
\end{equation}
Note the exchange of the signs of the $\varepsilon$ in the propagators. Similarly, for crossed box which we label by $u$ we get
\begin{equation}
 I_{\Box}^{u}=\frac{|\vec{\underline q}|^{D-6}}{4\hbar}\int \frac{d^D k}{(2\pi)^D} \frac{1}{(p_1 \cdot k+i \varepsilon)(p_4 \cdot k+i \varepsilon)k^2 (k+u_q)^2}
\end{equation}
or, equivalently,
\begin{equation}
 I_{\Box}^{u}=\frac{|\vec{\underline q}|^{D-6}}{4\hbar}\int \frac{d^D k}{(2\pi)^D} \frac{1}{(p_3 \cdot k-i \varepsilon)(p_2 \cdot k-i \varepsilon)k^2 (k+u_q)^2}
\end{equation}
Adding the boxes ($ I_{\Box}= I_{\Box}^{s}+ I_{\Box}^{u}$) the final
result takes the compact form
 \begin{multline}
 I_{\Box}=-\frac{|\vec{\underline q}|^{D-6}}{8 \hbar}\int \frac{d^D k}{(2\pi)^D} \frac{1}{k^2 (k+u_q)^2}\cr\times\left(\frac{1}{p_1 \cdot k+i\epsilon}-\frac{1}{p_3 \cdot k-i\epsilon}\right)\left(\frac{1}{p_2 \cdot k-i\epsilon}-\frac{1}{p_4 \cdot k+i\epsilon}\right).
\end{multline}
As shown in Appendix~\ref{sec:real}, the real part of the box integral
is given by  the unitarity cuts of the massive propagators
\begin{equation}
 \Re (I_{\Box}) =
-\frac{|\vec{\underline q}|^{D-6}}{2\hbar} \int \frac{d^D k}{(2\pi )^{D-2}} \frac{\delta((k+p_1)^2-m_1^2)\delta((k-p_2)^2-m_2^2)}{k^2 (k+u_q)^2},
\label{Ideltadelta}
\end{equation}
which exactly matches to all orders in $q^2$, the product of trees from the right
hand side of the one-loop unitarity relation~\eqref{firstUnitarity}
(recalling that the one-loop amplitude is $i$ times the box
contribution)
\begin{equation}
I_{\Box}^{\rm 1-cut} \equiv -2\Re(I_{\Box}) .  
\end{equation}
The integral~\eqref{Ideltadelta} evaluated in $D=4-2\epsilon$  reads
\begin{eqnarray}
 \Re(I_{\Box}) = 
 -\left(1- \frac{\hbar^2|\vec{\underline q}|^2 s}{4m_1^2m_2^2(\sigma^2-1)}\right)^{\epsilon}\frac{1}{8|\vec{\underline q}|^{2+2\epsilon}\hbar m_1 m_2 \sqrt{\sigma^2-1}} 
\frac{\Gamma(-\epsilon)^2
  \Gamma(1+\epsilon)}{(4\pi)^{1-\epsilon}
  \Gamma(-2\epsilon)}.
\label{Ideltadeltaeval}
\end{eqnarray}
This has precisely the form needed for being cancelled by the first unitarity subtraction of eq.~(\ref{Nmatrixelements}).

At two-loop order we need one more iteration of the tree-level
amplitude, which is the two-loop equivalent of the  one-cut (or two
velocity cuts) of the box
integral. The relevant two-cut (or four velocity cuts) two-loop integral is  in
$D=4-2\epsilon$ dimensions
\begin{multline}
 I_{\Box \Box}^{\rm 2-cut}=\hbar^3\int \frac{d^D l_1 d^D
   l_2}{(2\pi\hbar)^{2D-4}}
 \frac{\delta((p_1-l_1)^2-m_1^2)\delta((p_2+l_1)^2-m_2^2)}{l_1^2
    (l_1+l_2-q)^2}\cr
 \times  \frac{\delta((p_1+l_2-q)^2-m_1^2)\delta((p_2-l_2+q)^2-m_2^2)}{ l_2^2}.
\end{multline}
As we show in detail in Appendix~\ref{sec:real}, this integral can also be evaluated in the soft expansion. For both the computation of $q^2$-corrections to $\Re(I_{\Box})$
and $I_{\Box \Box}^{\rm 2-cut}$ our method is the one of velocity cuts introduced in ref.~\cite{Bjerrum-Bohr:2021din}. Velocity cuts can be viewed as the
leading-order parts of full unitarity cuts and they conveniently form the starting point for computations of $q^2$-corrections in the soft expansion. 

Keeping for now only the order in $q^2$ needed for our present purpose,
we find
\begin{multline}\label{e:realddbox}
  I_{\Box \Box}^{\rm 2-cut}
= - \left(1-\frac{\epsilon|\vec{\underline q}|^2 s}{3 m_1^2 m_2^2
    (\sigma^2-1)}\right) \frac{1}{ 16 m_1^2 m_2^2(\sigma^2-1) |\vec{\underline q}|^{2+4\epsilon}\hbar^3} \frac{\Gamma(-\epsilon)^3
  \Gamma(1+2\epsilon)}{(4 \pi)^{2-2\epsilon}\Gamma(-3\epsilon)}\cr+\mathcal O(|\vec{\underline q}|^{1-4\epsilon}).
\end{multline}
These equations, together with the results already provided in refs.~\cite{Bjerrum-Bohr:2021vuf,Bjerrum-Bohr:2021din} suffice to evaluate the needed matrix elements
of $\hat{N}_1$ and $\hat{N}_2$ for both maximal supergravity and Einstein gravity. This will be described in the next two subsections.

\subsection{Maximal supergravity}

In maximal supergravity the tree amplitude reads
\begin{equation}\label{e:M0}
\mathcal M_0(|\vec{\underline q}|,\sigma)= N_0(|\vec{\underline q}|,\sigma)=\frac{32 \pi
  G_Nm_1^2 m_2^2 \sigma^2}{|\vec{\underline q}|^2}
\end{equation}
and the one-loop amplitude evaluated in~\cite{Bjerrum-Bohr:2021vuf}
can  be rewritten as
\begin{multline}
\mathcal M_1(|\vec{\underline q}|,\sigma)={i\hbar\over2}\left(32 \pi  G_N m_1^2
  m_2^2\sigma^2\right)^2 I^{\rm 1-cut}_\Box\cr
+ \frac{32 \sqrt{\pi}  G_N^2 m_1^2 m_2^2(m_1+m_2)
  \sigma^4}{\sigma^2-1}\frac{(4\pi)^\epsilon
  \Gamma(\frac{1}{2}-\epsilon)^2
  \Gamma(\frac{1}{2}+\epsilon)}{ |\vec{\underline q}|^{1+2\epsilon}
  \Gamma(-2\epsilon)}\cr +\frac{16 (1+2\epsilon)G_N^2m_1^2 m_2^2
  \sigma^4(\sigma
  \arccosh(\sigma)-\sqrt{\sigma^2-1})}{(\sigma^2-1)^{\frac{3}{2}}}\frac{(4\pi)^\epsilon
  \Gamma(-\epsilon)^2 \Gamma(1+\epsilon)}{|\vec{\underline q}|^{2\epsilon} \Gamma(-2\epsilon)}\hbar+\mathcal O(|\vec{\underline q}|^{1+2\epsilon}).
\end{multline}
The first line is the square of the numerator of the tree amplitude
in~\eqref{e:M0} times the imaginary part of the one-loop amplitude
in~\eqref{Ideltadelta}.
This is just the unitarity subtraction given by one iteration of the
tree as shown in the one-loop computation above.
These amplitudes are given with the choice of helicity for the
external states made
in~\cite{Caron-Huot:2018ape,Parra-Martinez:2020dzs}. We refer to
section~3 of~\cite{Bjerrum-Bohr:2021vuf} for a discussion of helicity
dependence on the various part of the amplitude.
We thus immediately get from~\eqref{e:NtoT}
\begin{multline}
N_1(|\vec{\underline q}|,\sigma)=\frac{32 \sqrt{\pi}  G_N^2 m_1^2 m_2^2(m_1+m_2)
  \sigma^4}{\sigma^2-1}\frac{(4\pi)^\epsilon
  \Gamma(\frac{1}{2}-\epsilon)^2
  \Gamma(\frac{1}{2}+\epsilon)}{|\vec{\underline q}|^{1+2\epsilon}
  \Gamma(-2\epsilon)}\cr+\frac{16 (1+2\epsilon)G_N^2m_1^2 m_2^2
  \sigma^4(\sigma
  \arccosh(\sigma)-\sqrt{\sigma^2-1})}{(\sigma^2-1)^{\frac{3}{2}}}\frac{(4\pi)^\epsilon
  \Gamma(-\epsilon)^2 \Gamma(1+\epsilon)}{|\vec{\underline q}|^{2\epsilon} \Gamma(-2\epsilon)}\hbar+\mathcal O(|\vec{\underline q}|^{1+2\epsilon}).
\end{multline}
It is worthwhile to highlight one difference with the corresponding eikonal calculation here. Apart from exponentiating the amplitude in impact-parameter space, the eikonal method is
also based on an order-by-order separation into classical and quantum pieces conventionally parameterized in the $b$-space transform of the $S$-matrix as
\beq
\bar{S}(b) ~=~ \left(1 + 2\frac{i}{\hbar}\Delta(b)\right)e^{2i\delta(b)\over\hbar}.
\eeq
Where $\Delta(b)$ collects, to any given order in the expansion, those quantum terms that are not needed to ensure exponentiation. 
Indeed, already at one-loop order $\Delta(b)$ will contain $q^2$-corrections that are subtracted off in the above cancellations used to produce $N_1(|\vec{\underline q}|,\sigma)$. 
Thus, to that order, exponentiation of
the eikonal is also based on unitarity~\cite{Cristofoli:2020uzm} but only through the leading contribution from the unitarity relation. Here, we instead consistently subtract in
the exponent and thus include the $q^2$-corrections. In this way, all imaginary parts of the amplitude in momentum space are removed completely, rather than 
kept at the non-exponentiated level, as in the eikonal formalism. At this one-loop level, where we have kept the first quantum correction to illustrate our point,
this does not affect the calculation of the classical term but it shows explicitly how the imaginary quantum terms are removed. From 2-loop order and up these (real)
quantum subtractions of lower order can correct classical terms at higher
orders. A related phenomenon occurs in the eikonal formalism, but again: the details differ. 

\noindent
We next move to the two-loop amplitude in supergravity, including the
radiation reaction parts. Quoting from
ref.~\cite{Bjerrum-Bohr:2021vuf}, we can conveniently rewrite that
result as

\begin{multline}
\mathcal M_2(|\vec{\underline q}|,\sigma)={\hbar\over 6} (32\pi G_N m_1^2m_2^2\sigma^2)^3 I^{\rm 2-cut}_{\Box \Box}\cr
+\frac{64 i \sqrt{\pi} G_N^3 m_1^3 m_2^3 (m_1+m_2) \sigma^6
}{(\sigma^2-1)^{\frac{3}{2}}|\vec{\underline q}|^{1+4\epsilon}\hbar}\frac{(4\pi)^{2\epsilon}
  \Gamma(\frac{1}{2}-\epsilon)^2\Gamma(\frac{1}{2}+2\epsilon)
  \Gamma(-\epsilon)
  \Gamma(\frac{1}{2}-2\epsilon)}{\Gamma(\frac{1}{2}-3\epsilon)\Gamma(-2\epsilon)}\cr
+\frac{32 G_N^3 m_1^4 m_2^4 \sigma^4 (4\pi
  e^{-\gamma_E})^{2\epsilon}}{\pi}\Bigg(\frac{i \pi(1+2\epsilon) \sigma^2
  (\sigma \arccosh{\sigma}-\sqrt{\sigma^2-1})}{\epsilon^2
  |\vec{\underline q}|^{4\epsilon} m_1 m_2(\sigma^2-1)^{2}}\cr
- \frac{\pi^2 s \sigma^2}{6 \epsilon |\vec{\underline q}|^{4\epsilon}m_1^2 m_2^2
  (\sigma^2-1)^2} -\frac{\pi^2 \arccosh(\sigma)}{\epsilon
  |\vec{\underline q}|^{4\epsilon} m_1 m_2 \sqrt{\sigma^2-1}}\cr
-\frac{i\pi(1+i\pi\epsilon)}{2
  \epsilon^2 |\vec{\underline q}|^{4\epsilon}m_1 m_2
  (\sigma^2-1)^2}
\Big((1+2\epsilon)\sigma^2\sqrt{\sigma^2-1}
+\sigma(\sigma^2-2)\arccosh(\sigma)+\epsilon((\sigma^2-1)^{\frac{3}{2}}\cr-\sigma(\sigma^2-2))\arccosh^2(\sigma)
-\epsilon \sigma(\sigma^2-2)\Li_2 \big(2-2\sigma(\sigma+\sqrt{\sigma^2-1})\big) \Big)\Bigg) +\mathcal O(\hbar).
\end{multline}
In this form we see that the first three lines are eliminated by  the relations of eq.~(\ref{Nmatrixelements}) with tree and one-loop terms. The imaginary part of the radiation
reaction part is subtracted by the corresponding two-to-three particle cut of eq.~(\ref{Nmatrixelements}) as follows from the calculation of ref.~\cite{DiVecchia:2021bdo}.
We thus get
\begin{multline}
N_2(|\vec{\underline q}|,\sigma)=\frac{32 \pi G_N^3 m_1^3 m_2^3 \sigma^4
  \left(4\pi e^{-\gamma_E}\right)^{2\epsilon}}{\epsilon
    |\vec{\underline q}|^{4\epsilon} \sqrt{\sigma^2-1}}
  \Bigg(- \frac{ s   \sigma^2}{6 m_1 m_2 (\sigma^2-1)^{\frac{3}{2}}}
  -\arccosh(\sigma)\\
  + \Big(\frac{1}{4(\sigma^2-1)}\Big)^\epsilon
  \Big(\frac{\sigma(\sigma^2-2)\arccosh(\sigma)}{(\sigma^2-1)^{3\over2}}+\frac{\sigma^2}{\sigma^2-1}
 \Big)\Bigg)+\mathcal O(\hbar). 
\end{multline}
Up to two loop classical order and keeping only the leading terms in $\epsilon$ we finally have
\begin{multline}
N(|\vec{\underline q}|,\sigma)=\frac{32 \pi G_Nm_1^2 m_2^2
  \sigma^2}{|\vec{\underline q}|^2}+\frac{32 \pi G_N^3 m_1^3 m_2^3 \sigma^4
}{\epsilon |\vec{\underline q}|^{4\epsilon} \sqrt{\sigma^2-1}}\Bigg(- \frac{ s
  \sigma^2}{6 m_1 m_2 (\sigma^2-1)^{\frac{3}{2}}} -\arccosh(\sigma)\cr
+\Big(\frac{1}{4(\sigma^2-1)}\Big)^\epsilon \Big(\frac{\sigma(\sigma^2-2)\arccosh(\sigma)}{(\sigma^2-1)^{3\over2}}+\frac{\sigma^2}{\sigma^2-1} \Big)\Bigg) +\mathcal O(\hbar).
\end{multline}
Following the prescription of ref.~\cite{Bern:2021dqo} and thus defining a Fourier transform of a function $f(s,q^2)$ by
\begin{equation}
\bar{f}(s,b^2)=\frac{1}{4m_1 m_2 \sqrt{\sigma^2-1}}\int \frac{d^{D-2}
  \underline q}{(2\pi)^{D-2}}f(s,q^2) e^{-i b\cdot \underline q}
\end{equation}
we go to impact-parameter space with $\vec{p}\cdot \vec{b}=0$ and 
angular momentum $J= |\vec{p}|  |\vec{b}|$. This gives, again keeping
only needed terms of $\epsilon$,
\begin{multline}
\bar{N}(b,\sigma)=\frac{2G_Nm_1 m_2
  \sigma^2\Gamma(-\epsilon)}{\sqrt{\sigma^2-1}} (\pi
|\vec{b}|^2)^{\epsilon}+\frac{16 \pi G_N^3 m_1^2 m_2^2
  \sigma^4}{\sigma^2-1}\Bigg(- \frac{ s \sigma^2}{6 m_1 m_2
  (\sigma^2-1)^{\frac{3}{2}}} -\arccosh(\sigma)\cr
+\Big(\frac{1}{4(\sigma^2-1)}\Big)^\epsilon
\Big(\frac{\sigma(\sigma^2-2)\arccosh(\sigma)}{(\sigma^2-1)^{3\over2}}+\frac{\sigma^2}{\sigma^2-1}
\Big)\Bigg){1\over (\pi |\vec{b}|^2)^{1-3\epsilon}}+\mathcal O(\hbar)
\end{multline}
or, in terms of $J$ and ignoring factors of $p^{\epsilon}$ which become unity in the $\epsilon \to 0$ limit,

\begin{multline}
\bar{N}(J,\sigma)=\frac{2G_Nm_1 m_2
  \sigma^2\Gamma(-\epsilon)}{\sqrt{\sigma^2-1}} (\pi
J^2)^{\epsilon}+\frac{16 \pi G_N^3 m_1^4 m_2^4 \sigma^4}{s}\Bigg(- \frac{ s
  \sigma^2}{6 m_1 m_2 (\sigma^2-1)^{\frac{3}{2}}} -\arccosh(\sigma)\cr
+\Big(\frac{1}{4(\sigma^2-1)}\Big)^\epsilon
\Big(\frac{\sigma(\sigma^2-2)\arccosh(\sigma)}{(\sigma^2-1)^{3\over2}}+\frac{\sigma^2}{\sigma^2-1}
\Big)\Bigg){1\over (\pi J^2)^{1-3\epsilon}}+\mathcal O(\hbar).
\end{multline}
Assuming that this coincides with the interacting part of the radial action, we get
\begin{multline}
\chi=-\frac{\partial}{\partial J}\lim_{\hbar\to0}\bar{N}(J,\sigma)=\frac{4G_Nm_1 m_2
  \sigma^2}{\sqrt{\sigma^2-1}}\frac{1}{J} \cr+\frac{32 G_N^3 m_1^4
  m_2^4 \sigma^4}{s}\Bigg(- \frac{ s \sigma^2}{6 m_1 m_2
  (\sigma^2-1)^{\frac{3}{2}}} -\arccosh(\sigma)+
 {d\over d\sigma}\left( \sigma^2
  \arccosh(\sigma)\over \sqrt{\sigma^2-1}\right)\Bigg)\frac{1}{J^3}
\end{multline}
which we indeed recognize as the angle at third Post-Minkowskian order in maximal supergravity~\cite{DiVecchia:2020ymx,DiVecchia:2021bdo,Bjerrum-Bohr:2021vuf}. It includes
all terms, and hence also radiation reaction pieces.

\subsection{General relativity}

Next, we turn to Einstein gravity. We will be able to recycle much of what was used above plus add the needed new features from refs. 
\cite{Bjerrum-Bohr:2021vuf,Bjerrum-Bohr:2021din}. 

The tree and one-loop amplitudes now read
\begin{equation}
\mathcal M_0(|\vec{\underline q}|,\sigma)= N_0(|\vec{\underline q}|,\sigma)=\frac{16 \pi
  G_Nm_1^2 m_2^2(2\sigma^2-1)}{|\vec{\underline q}|^2}+\mathcal O(\hbar)
\end{equation}
and
\begin{multline}
\mathcal M_1(|\vec{\underline q}|,\sigma)={i\hbar\over2} (16\pi G_N m_1^2m_2^2(2\sigma^2-1) )^2 I_{\Box}^{\rm 1-cut}\\+ \frac{3 \pi^2  G_N^2 m_1^2 m_2^2(m_1+m_2) (5\sigma^2-1)(4\pi e^{-\gamma_E})^\epsilon}{|\vec{\underline q}|^{1+2\epsilon}} \\-\frac{4G_N^2m_1^2 m_2^2(4\pi e^{-\gamma_E})^{\epsilon}\hbar}{\epsilon |\vec{\underline q}|^{2\epsilon}} \Big(\frac{2(2\sigma^2-1)(7-6\sigma^2)\arccosh(\sigma)}{(\sigma^2-1)^{\frac{3}{2}}}+\frac{1 - 49 \sigma^2 + 18 \sigma^4}{15(\sigma^2-1)}\Big)+\mathcal O(|\vec{\underline q}|^{1+2\epsilon})
\end{multline}
where for the classical and quantum terms this is valid up to leading pieces in $\epsilon$. For the one-loop piece we have rewritten the result in the same manner
as for maximal supergravity above. We notice that the first line is cancelled by the unitarity relation based on the iteration of the tree as dictated by 
eq.~(\ref{Nmatrixelements}), leaving us with

\begin{multline}
N_1(|\vec{\underline q}|,\sigma)= \frac{3 \pi^2  G_N^2 m_1^2
  m_2^2(m_1+m_2) (5\sigma^2-1)(4\pi
  e^{-\gamma_E})^\epsilon}{|\vec{\underline q}|^{1+2\epsilon}} \cr
-\frac{4G_N^2m_1^2 m_2^2(4\pi e^{-\gamma_E})^{\epsilon}\hbar}{\epsilon |\vec{\underline q}|^{2\epsilon}} \Big(\frac{2(2\sigma^2-1)(7-6\sigma^2)\arccosh(\sigma)}{(\sigma^2-1)^{\frac{3}{2}}}+\frac{1 - 49 \sigma^2 + 18 \sigma^4}{15(\sigma^2-1)}\Big)+\mathcal O(|\vec{\underline q}|^{1+2\epsilon}).
\end{multline}
We next turn to the full two-loop amplitude in Einstein gravity~\cite{Bjerrum-Bohr:2021din} which to the needed order in $q^2$  can be rewritten as
\begin{multline}
\mathcal M_2(|\vec{\underline
  q}|,\sigma)={\hbar\over6}(16 \pi G_N m_1^2 m_2^2
  (2\sigma^2-1))^3  I_{\Box \Box}^{\rm 2-cut}\cr
+\frac{6i \pi^2 G_N^3
  (m_1+m_2) m_1^3 m_2^3(2\sigma^2-1)(1-5\sigma^2)(4\pi
  e^{-\gamma_E})^{2\epsilon}}{\epsilon
  \sqrt{\sigma^2-1}|\vec{\underline q}|^{1+4\epsilon}}\cr
+\frac{2 \pi
  G_N^3(4\pi e^{-\gamma_E})^{2\epsilon} m_1^2 m_2^2\hbar}{ \epsilon
  |\vec{\underline q}|^{4\epsilon}}\Bigg(\frac{im_1
  m_2(2\sigma^2-1)}{\pi \epsilon
  (\sigma^2-1)^{\frac{3}{2}}}\Big(\frac{1-49\sigma^2+18
  \sigma^4}{15}-\frac{2\sigma(7-20 \sigma^2+12
  \sigma^4)\arccosh(\sigma)}{\sqrt{\sigma^2-1}}\Big)\cr
+\frac{s \left(64 \sigma ^6-120 \sigma ^4+60 \sigma
   ^2-5\right)}{3\left(\sigma ^2-1\right)^2}-{4\over3} m_1 m_2 \sigma  \left(14 \sigma
   ^2+25\right)\cr
+\frac{4 m_1 m_2(3+12
  \sigma^2-4\sigma^4)\arccosh(\sigma)}{\sqrt{\sigma^2-1}} \cr-\frac{2i
  m_1 m_2(2\sigma^2-1)^2}{\pi \epsilon \sqrt{\sigma^2-1}} {\frac{1+i\pi
  \epsilon}{(4(\sigma^2-1))^{\epsilon}}}\bigg(-\frac{11}{3}+\frac{d}{d\sigma}
\Big(\frac{(2\sigma^2-1)\arccosh(\sigma)}{\sqrt{\sigma^2-1}} \Big)
\bigg)\Bigg)+\mathcal O(\hbar).
\end{multline}
So that the first three lines are eliminated by the subtractions of tree and one-loop terms as dictated by eq.~(\ref{Nmatrixelements}). The imaginary part of the radiation
reaction term is cancelled by the corresponding two-to-three particle cut of eq.~(\ref{Nmatrixelements}) as again follows from the calculation of ref.~\cite{DiVecchia:2021bdo}. We
thus find
\begin{multline}
N_2(|\vec{\underline q}|,\sigma)=\frac{2 \pi G_N^3(4\pi
  e^{-\gamma_E})^{2\epsilon} m_1^2 m_2^2 }{\epsilon |\vec{\underline q}|^{4\epsilon}}\Bigg(\frac{s \left(64 \sigma ^6-120 \sigma ^4+60 \sigma
   ^2-5\right)}{3\left(\sigma ^2-1\right)^2}\cr-{4\over3} m_1 m_2 \sigma  \left(14 \sigma
   ^2+25\right)+\frac{4 m_1 m_2(3+12
   \sigma^2-4\sigma^4)\arccosh(\sigma)}{\sqrt{\sigma^2-1}}
 \\+\frac{2m_1 m_2(2\sigma^2-1)^2}{
   \sqrt{\sigma^2-1}}\bigg(-\frac{11}{3}+\frac{d}{d\sigma}
 \Big(\frac{(2\sigma^2-1)\arccosh(\sigma)}{\sqrt{\sigma^2-1}} \Big)
 \bigg)\Bigg)+\mathcal O(\hbar).
\end{multline}
Up to two loop classical order and again keeping only the leading real terms in $\epsilon$ we thus arrive at
\begin{multline}
N(|\vec{\underline q}|,\sigma)=\frac{16 \pi G_Nm_1^2
  m_2^2(2\sigma^2-1)}{|\vec{\underline q}|^2}+\frac{3 \pi^2  G_N^2
  m_1^2 m_2^2(m_1+m_2) (5\sigma^2-1)}{|\vec{\underline
    q}|^{1+2\epsilon}}\cr
+\frac{2 \pi G_N^3(4\pi
  e^{-\gamma_E})^{2\epsilon} m_1^2 m_2^2 }{\epsilon |\vec{\underline
    q}|^{4\epsilon}}\Bigg(\frac{s \left(64 \sigma ^6-120 \sigma ^4+60 \sigma
   ^2-5\right)}{3\left(\sigma ^2-1\right)^2}\cr-{4\over3} m_1 m_2 \sigma  \left(14 \sigma
   ^2+25\right)+\frac{4 m_1 m_2(3+12
  \sigma^2-4\sigma^4)\arccosh(\sigma)}{\sqrt{\sigma^2-1}} \cr
+\frac{2 m_1 m_2(2\sigma^2-1)^2}{\sqrt{\sigma^2-1}}
\bigg(-\frac{11}{3}+\frac{d}{d\sigma}
\Big(\frac{(2\sigma^2-1)\arccosh(\sigma)}{\sqrt{\sigma^2-1}} \Big)
\bigg)\Bigg)+\mathcal O(\hbar)
\end{multline}
which after the Fourier transform to impact-parameter space becomes
\begin{multline}
\bar{N}(b,\sigma)=\frac{G_Nm_1 m_2
  (2\sigma^2-1)\Gamma(-\epsilon)}{\sqrt{\sigma^2-1}} (\pi
|\vec{b}|^2)^{\epsilon}+\frac{3 \pi^{\frac{3}{2}}  G_N^2 m_1
  m_2(m_1+m_2) (5\sigma^2-1)}{4\sqrt{\sigma^2-1} (\pi
|\vec{b}|^2)^{\frac{1}{2}-2\epsilon}}\cr
+\frac{ \pi G_N^3 m_1 m_2
}{\sqrt{\sigma^2-1} (\pi |\vec{b}|^2)^{1-3\epsilon}}\Bigg(\frac{s \left(64 \sigma ^6-120 \sigma ^4+60 \sigma
   ^2-5\right)}{3\left(\sigma ^2-1\right)^2}\cr-{4\over3} m_1 m_2 \sigma  \left(14 \sigma
   ^2+25\right)+\frac{4 m_1 m_2(3+12
  \sigma^2-4\sigma^4)\arccosh(\sigma)}{\sqrt{\sigma^2-1}} \cr
+\frac{2 m_1 m_2(2\sigma^2-1)^2}{\sqrt{\sigma^2-1}}
\bigg(-\frac{11}{3}+\frac{d}{d\sigma}
\Big(\frac{(2\sigma^2-1)\arccosh(\sigma)}{\sqrt{\sigma^2-1}} \Big)
\bigg)\Bigg)+\mathcal O(\hbar)
\end{multline}
or, in terms of angular momentum $J$,
\begin{multline}
\bar{N}(J,\sigma)=\frac{G_Nm_1 m_2
  (2\sigma^2-1)\Gamma(-\epsilon)}{\sqrt{\sigma^2-1}}
J^{2\epsilon}+\frac{3 \pi G_N^2 m_1^2 m_2^2(m_1+m_2) (5\sigma^2-1)}{4
  \sqrt{s}}\frac{1}{J}\cr
+\frac{G_N^3 m_1^3 m_2^3\sqrt{\sigma^2-1}
}{s}\Bigg(\frac{s \left(64 \sigma ^6-120 \sigma ^4+60 \sigma
   ^2-5\right)}{3\left(\sigma ^2-1\right)^2}\cr-{4\over3} m_1 m_2 \sigma  \left(14 \sigma
   ^2+25\right)+\frac{4 m_1 m_2(3+12
   \sigma^2-4\sigma^4)\arccosh(\sigma)}{\sqrt{\sigma^2-1}} \\+\frac{2
   m_1 m_2(2\sigma^2-1)^2}{\sqrt{\sigma^2-1}}
 \bigg(-\frac{11}{3}+\frac{d}{d\sigma}
 \Big(\frac{(2\sigma^2-1)\arccosh(\sigma)}{\sqrt{\sigma^2-1}} \Big)
 \bigg)\Bigg)\frac{1}{J^2}+\mathcal O(\hbar).
\end{multline}
Taking this to be the interacting part of the radial action to third Post-Minkowskian order, we obtain the scattering angle
\begin{multline}
\chi=-\frac{\partial}{\partial J}\lim_{\hbar\to0}\bar{N}(J,\sigma)=\frac{2G_Nm_1 m_2 (2\sigma^2-1)}{\sqrt{\sigma^2-1}} \frac{1}{J}+\frac{3 \pi G_N^2 m_1^2 m_2^2(m_1+m_2) (5\sigma^2-1)}{4 \sqrt{s}}\frac{1}{J^2}\\+\frac{2G_N^3 m_1^3 m_2^3\sqrt{\sigma^2-1} }{s}\Bigg(\frac{s \left(64 \sigma ^6-120 \sigma ^4+60 \sigma
   ^2-5\right)}{3\left(\sigma ^2-1\right)^2}\cr-{4\over3} m_1 m_2 \sigma  \left(14 \sigma
   ^2+25\right)+\frac{4 m_1 m_2(3+12 \sigma^2-4\sigma^4)\arccosh(\sigma)}{\sqrt{\sigma^2-1}} \\+\frac{2 m_1 m_2(2\sigma^2-1)^2}{\sqrt{\sigma^2-1}} \bigg(-\frac{11}{3}+\frac{d}{d\sigma} \Big(\frac{(2\sigma^2-1)\arccosh(\sigma)}{\sqrt{\sigma^2-1}} \Big) \bigg)\Bigg)\frac{1}{J^3} ~,
\end{multline}
which agrees with the literature~\cite{Bern:2019nnu,Antonelli:2019ytb}, including the radiation reaction terms~\cite{DiVecchia:2020ymx,Damour:2020tta,Bjerrum-Bohr:2021din}.

\section{Kinematics in isotropic coordinates}

So far, the computation of the scattering angle from the amplitude has entirely bypassed the notion of a potential $V$; only the radial action, an indirect function of
the potential, played a role. However, there is more information in the potential $V$ itself, even if it by construction refers to specific coordinates. It may therefore
be useful to see how such an effective potential can be extracted from
the scattering angle. Our starting point for this is the relativistic Salpeter equation of two-body
scattering in the center of mass frame,
\beq
E ~=~ \sqrt{p^2 + m_1^2} + \sqrt{p^2 + m_2^2} + V(p,r) ~.
\label{Energy}
\eeq
The operator version of this equation together with the Lippmann-Schwinger equation~\cite{Cristofoli:2019neg} allows us to relate the potential $V(p,r)$ to the
Fourier transform of the scattering amplitude. However, it is not necessary to introduce this additional step if we already know the scattering angle up to the given
order in $G_N$. In isotropic coordinates we can always solve the energy equation~(\ref{Energy}) in terms of $p^2$,
\begin{equation}
p^2 ~=~ p_{\infty}^2  -  V_{\rm eff}(r) ~,
\label{kinematics}
\end{equation}
where $p_\infty^2=m_1^2m_2^2(\sigma^2-1)/s$ and,  without  loss of generality we can parametrize (in $D=4$ dimensions) 
\beq
V_{\rm eff}(r) ~=~ -\sum_{n=1}^{\infty} \frac{G_N^nf_n(E)}{r^n}
\label{Veff}
\eeq 
where the coefficients must be extracted from the amplitude\footnote{The case of general $D$ is discussed in ref.~\cite{Cristofoli:2020uzm}.}. One of the surprising
results of the amplitude approach is that $V_{\rm eff}(r)$ is directly related to the classical part of the amplitude as it derives either from the effective field theory matching
\cite{Bern:2019crd} or from the Born subtractions~\cite{Kalin:2019rwq,Bjerrum-Bohr:2019kec}. In fact, the kinematical relation~(\ref{kinematics}) can be taken
as a new and equally good quantum mechanical Hamiltonian operator, a result anticipated by Damour~\cite{Damour:2017zjx} before these explicit amplitude computations.

The scattering angle based on the kinematical relation~(\ref{kinematics}) has been derived to all orders and we quote the first few orders from Table 1 of
ref.~\cite{Bjerrum-Bohr:2019kec},
\begin{align}
\chi_{1PM} &= f_1, \cr
\chi_{2PM} &= {\pi p_{\infty}\over2} f_2, \cr
\chi_{3PM} &=  2p_{\infty}^4f_3 + p_{\infty}^2f_1f_2 - {f_1^3\over 12} ~.
\end{align}
Comparing with the scattering angle computed to this order by the eikonal method (and reproduced here, using the new method) we can recursively solve for the
unknown $f_i$-coefficients. 

Reminding the reader that $s = m_1^2 + m_2^2 + 2m_1m_2\sigma$ also has $\sigma$-dependence, we find
\begin{align}\label{e:f1PM}
f_1 &= {2m_1^2m_2^2(2\sigma^2 -1)\over\sqrt{s}} ,\\
\label{e:f2PM} f_2 &= \frac{3 m_1^2 m_2^2 \left(5 \sigma ^2-1\right) (m_1+m_2)}{2
   \sqrt{s}},\\
\label{e:f3PM} f_3 &=-\frac{m_1^2 m_2^2}{2 \left(\sigma
        ^2-1\right)}\,\Big(3 \left(2 \sigma ^2-1\right) \left(5
        \sigma ^2-1\right)
        (m_1+m_2) 
        -4 \left(12 \sigma ^4-10 \sigma
          ^2+1\right) \sqrt{s} \Big)\cr
       &   -\frac{2m_1^3 m_2^3}{3 \sqrt{s} }\,\left(2\sigma (14 \sigma ^2+25)+\frac{6 \left(4 \sigma ^4-12 \sigma ^2-3\right)}{\sqrt{\sigma ^2-1}}\arccosh(\sigma
          )\right)\cr
\nonumber           &  +\frac{2 m_1^3 m_2^3 \left(1-2 \sigma ^2\right)^2}{3 \sqrt{s} \left(\sigma ^2-1\right)^2}\left((8-5 \sigma ^2)\sqrt{\sigma^2 - 1}+(6 \sigma ^3-9 \sigma ) \arccosh
   (\sigma )\right),\\ 
\end{align}
including radiation-reaction contributions in the last line of
$f_3$. We stress again that while these $f_i$-coefficients reproduce
the scattering angle up to third post-Minkowskian order including the contributions
from radiation-reaction terms, they also, through eq.~(\ref{Veff}), provide us with the kinematical relation~(\ref{kinematics}) in isotropic coordinates.

Comparing to the potential of a probe small mass $m$ in the
Schwarzschild background of large mass $M\gg m$ we have~\cite{Damour:2017zjx}
\begin{align}
  \label{e:fprobe}
  f_1^{\rm probe}&=2 m^2  M (2\sigma^2-1),\cr
   f_2^{\rm probe}&={3\over2}m^2M^2(5\sigma^2-1),\cr
   f_3^{\rm probe}&=         {m^2M^3\over 2}(18\sigma^2-1)    ~,                    
\end{align}
a well known result when comparing with the result of the potential region only. Because radiation reaction terms vanish in the probe limit, this 
is indeed unchanged here.

\section{Conclusions}

In an attempt to improve on the systematic expansion of the eikonal formalism we have
instead explored an alternative idea recently suggested by Bern et al.~\cite{Bern:2021dqo} and which we find is linked to the closely related WKB approximation.
Using an exponential representation of the $S$-matrix, we systematically relate matrix elements of the operator in the exponential $\hat{N}$ to ordinary Born amplitudes minus
pieces provided by unitarity cuts. Crucially, we must now relate this object to the radial action. 
We  do this by a Fourier transform into impact parameter space and we have checked up to third Post-Minkowskian order that this method, combined with the above transformation to impact-parameter space, 
works for both maximal supergravity and Einstein gravity. It reproduces the scattering angles to that order and it is not limited to what is known as the potential region
of the loop amplitudes. Instead, we sum all classical contributions and thus include also radiation reaction pieces. The simplicity of this method seems very appealing and
suggests that it may be used to streamline Post-Minkowskian amplitudes in gravity by means of a diagrammatic technique that systematically avoids the evaluation of the cut
diagrams that must be subtracted, but simply discards them at the integrand level.

In practice, we need only evaluate matrix elements in the soft $q^2$-expansion. This means that we expand genuine unitarity cuts around the velocity cuts introduced
recently~\cite{Bjerrum-Bohr:2021vuf,Bjerrum-Bohr:2021din}. These velocity cuts seem to provide the most natural way to organize amplitude calculations in the soft expansion.

We have finally pointed out that there is no obstacle towards
obtaining the potential $V_{\rm eff}(r)$ 
from the scattering angles computed by this method. Iteratively, 
coefficients of the effective potential in isotropic coordinates 
follow from the angles and the result is unique. It is also not limited to the result of just the potential region of the amplitudes. This
thereby gives the kinematical relation between momenta and coordinates in isotropic gauge. As a simple illustration, one can from this predict infinite series of terms from
lower-order pieces. For instance, keeping only the leading $f_1$-term of the effective potential there is nothing to prevent one from summing the whole series to
obtain the standard Newtonian deflection angle. If one were to include also the $f_2$-term one would get the exact analytical result corresponding to the $f_1\!-\!f_2$-theory
computed in~\cite{Cristofoli:2019neg}. Of course, in the context of general relativity it is not meaningful to compute only a part of the higher order terms.

\vspace{1cm}
\noindent
\acknowledgments We thank Zvi Bern and Radu Roiban for
useful comments on the manuscript.  We also acknowledge interesting discussions with participants at the GGI workshop "Gravitational scattering, inspiral, and radiation", April-May, 2021.
The research of P.V. has received funding from the ANR grant
``Amplitudes'' ANR-17- CE31-0001-01, and the ANR grant ``SMAGP''
ANR-20-CE40-0026-01 and is partially supported by Laboratory of Mirror
Symmetry NRU HSE, RF Government grant, ag. No 14.641.31.0001. P.V. is
grateful to the I.H.E.S. for the use of their computer resources. The
work of P.H.D. was supported in part by DFF grant 0135-00089A.

\appendix
\section{Details of the one and two loop box calculations}\label{sec:real}

In this appendix we first provide some details on the derivation of the cut part of
the one-loop box integral in~\eqref{Ideltadelta} and two-cut part of
the double-box
integral in~\eqref{e:realddbox}. We show explicitly that the real part is
given by the unitarity cut of the massive propagators
in~\eqref{e:boxcut} and we evaluate it to all order in
$\underline q$ in~\eqref{e:boxrealfinal}.

\subsection{The cut part of the box integral}
The box integral is defined as
\begin{equation}
I_{\Box}^{s}=-\frac{1}{4}\int \frac{d^D k}{(2\pi \hbar)^D} \frac{\hbar^4}{(p_1 \cdot k+i \varepsilon)(p_2 \cdot k-i \varepsilon)k^2 (k+q)^2} ~.
\end{equation}
In dimensional regularization we can neglect the $k^2$-terms in the massive
propagators as they will, after cancelling a massless $k^2$-propagator only give rise to tadpoles in the soft expansion, and will hence be set to zero.
To perform the soft expansion in powers of $q= \hbar |\vec{\underline
  q} |  u_q$ where $u_q$ is a unit space-like vector ($u_q^2=-1$), we
make the change of variable $k \rightarrow \hbar |\vec{q}| k$ to get
\begin{equation}
I_{\Box}^{s}=-\frac{|\vec{\underline q}|^{D-6}}{4 \hbar^2}\int \frac{d^D k}{(2\pi)^D} \frac{1}{(p_1 \cdot k+i \varepsilon)(p_2 \cdot k-i \varepsilon)k^2 (k+u_q)^2}.
\end{equation}
We note that the integral can equivalently be written in terms of $p_3$ and $p_4$ as 
\begin{equation}
I_{\Box}^{s}=-\frac{|\vec{\underline q}|^{D-6}}{4 \hbar^2} \int \frac{d^D k}{(2\pi)^D} \frac{1}{(p_3 \cdot k-i \varepsilon)(p_4 \cdot k+i \varepsilon)k^2 (k+u_q)^2}
\end{equation}
with the important change of sign of signs of the $i\varepsilon$ term
in the propagators.

Similarly, in the $u$ channel corresponding to the crossed box integral, we have

\begin{align}
I_{\Box}^{u}&=\frac{|\vec{\underline q}|^{D-6}}{4 \hbar^2}\int \frac{d^D
              k}{(2\pi)^D} \frac{1}{(p_1 \cdot k+i \varepsilon)(p_4
              \cdot k+i \varepsilon)k^2 (k+u_q)^2},\cr
              &=\frac{|\vec{\underline q}|^{D-6}}{4 \hbar^2}\int \frac{d^D k}{(2\pi)^D} \frac{1}{(p_3 \cdot k-i \varepsilon)(p_2 \cdot k-i \varepsilon)k^2 (k+u_q)^2}.
\end{align}
The sum of these box contributions
$I_{\Box}=I_{\Box}^{s}+I_{\Box}^{u}$ takes the form
 \begin{multline}
I_{\Box}=-\frac{|\vec{\underline q}|^{D-6}}{8 \hbar^2}\int \frac{d^D k}{(2\pi)^D}\frac{1}{k^2 (k+u_q)^2}\cr\times \left(\frac{1}{p_1 \cdot k+i\varepsilon}-\frac{1}{p_3 \cdot k-i\varepsilon}\right)\left(\frac{1}{p_2 \cdot k-i\varepsilon}-\frac{1}{p_4 \cdot k+i\varepsilon}\right).
\end{multline}
Using the variables $p_1=\bar p_1+\hbar {\underline q\over2}$ and
$p_2=\bar p_2-\hbar {\underline q\over2}$ we have
 \begin{multline}
 I_{\Box}=-\frac{|\vec{\underline q}|^{D-6}}{8 \hbar^2}\int \frac{d^D
   k}{(2\pi)^D} \frac{1}{k^2 (k+u_q)^2}\cr
 \times\left(\frac{1}{\bar{p}_1 \cdot k+\frac{\hbar
       |\vec{\underline q}| u_q \cdot
       k}{2}+i\varepsilon}-\frac{1}{\bar{p_1} \cdot k-\frac{\hbar
       |\vec{\underline q}| u_q \cdot k}{2}-i\varepsilon}\right)\cr
\times\left(\frac{1}{\bar{p}_2 \cdot k-\frac{\hbar |\vec{\underline
         q}| u_q \cdot k}{2}-i\varepsilon}-\frac{1}{\bar{p}_2 \cdot
     k+\frac{\hbar |\vec{\underline q}| u_q \cdot
       k}{2}+i\varepsilon}\right).
\end{multline}
The soft expansion for small $\hbar |\underline q|$ reads
 \begin{multline}\label{e:Iboxsoft}
I_{\Box}=-\frac{|\vec{\underline q}|^{D-6}}{8
  \hbar^2}\sum_{n_1=0}^{\infty} \sum_{n_2=0}^{\infty}\int \frac{d^D
  k}{(2\pi)^D} \frac{1}{k^2 (k+u_q)^2} \left(\frac{\hbar|\vec{\underline
        q}|}{2}\, u_q\cdot k\right)^{n_1+n_2}\cr
\times\left(\frac{(-1)^{n_1}}{(\bar{p}_1 \cdot
    k+i\varepsilon)^{n_1+1}}-\frac{1}{(\bar{p}_1 \cdot
    k-i\varepsilon)^{n_1+1}}\right)
\left(\frac{1}{(\bar{p}_2 \cdot k-i\varepsilon)^{n_2+1}}-\frac{(-1)^{n_2}}{(\bar{p}_2 \cdot k+i\varepsilon)^{n_2+1}}\right).
\end{multline}
We now rewrite $2u_q\cdot k= (k+u_q)^2-k^2-u_q^2=(k+u_q)^2-k^2+1$. Since, as explained above,
we can neglect the tadpoles in dimensional regularization,  we can
replace  $u_q\cdot k$ by $1/2$ in the previous expression, and
reduce each integral in a basis of four master integrals of the
scalar box, the scalar triangles and the bubble (but we neglect
tadpoles as usual):
\begin{multline}
\int \frac{d^D k}{(2\pi)^D} \frac{1}{(\bar{p}_1 \cdot
  k)^{n_1+1}(\bar{p}_2 \cdot k)^{n_2+1}k^2 (k+u_q)^2}\cr
= \mathcal B_{n_1,n_2}(m_1,m_2,\hbar|\vec{\underline q}|,\sigma)\int
\frac{d^D k}{(2\pi)^D} \frac{1}{(\bar{p}_1 \cdot k)(\bar{p}_2 \cdot
  k)k^2 (k+u_q)^2}\cr
+\mathcal T^1_{n_1,n_2}(m_1,m_2,\hbar|\vec{\underline q}|,\sigma)\int
\frac{d^D k}{(2\pi)^D} \frac{1}{(\bar{p}_1 \cdot k)k^2 (k+u_q)^2}\cr
+\mathcal T^2_{n_1,n_2}(m_1,m_2,\hbar|\vec{\underline q}|,\sigma)\int
\frac{d^D k}{(2\pi)^D} \frac{1}{(\bar{p}_2 \cdot k)k^2 (k+u_q)^2}\cr
+\mathcal C_{n_1,n_2}(m_1,m_2,\hbar |\vec{\underline q}|,\sigma)\int \frac{d^D k}{(2\pi)^D} \frac{1}{k^2 (k+u_q)^2},
\end{multline}
where $\mathcal B$, $\mathcal T^1$, $\mathcal T^2$ and $\mathcal C$
are real rational functions of $m_1$, $m_2$, $\hbar|\vec{\underline
  q}|$ and $\sigma$. Since the bubble integral is purely imaginary this term cannot
contribute to the real part of the integral.

\bigskip
Using {\tt LiteRed}~\cite{Lee:2013mka}, we observe\footnote{This has
  been checked to high order in $|\underline q|^2$.} that for all
$n_1$ and $n_2$ we have  $\mathcal T^i_{2n_1,2n_2}=\mathcal
T^i_{2n_1+1,2n_2+1}=0$ with $i=1,2$. This implies that  the triangle
master integrals contribute only  when $n_1$ and $n_2$ are of
different parity. In that case, the sum of $s$ and $u$ channel in the
box in~\eqref{e:Iboxsoft} implies that the integral has one delta function, leaving an integral of the form
\begin{equation}
i \int \frac{d^{D-1} \vec{k}}{(2\pi)^{D-1}} \frac{1}{\vec{k}^2 (\vec{k}+\vec{u_q})^2}
\end{equation}
which is obviously imaginary.

We therefore conclude that only the box master integral contributes to
the real part of $I_{\Box}$.
This box master integral reads
 \begin{multline}
I_{\Box}|_{\rm box}=-\frac{|\vec{\underline q}|^{D-6}}{8
  \hbar^2}\sum_{n_1=0}^{\infty} \sum_{n_2=0}^{\infty} \left(\frac{\hbar
  |\vec{\underline q}|}{4}\right)^{n_1+n_2} \mathcal B_{n_1,n_2} \cr
\times\int \frac{d^D k}{(2\pi)^D}\frac{1}{k^2 (k+u_q)^2} \left(\frac{(-1)^{n_1}}{\bar{p}_1 \cdot k+i\varepsilon}-\frac{1}{\bar{p}_1 \cdot k-i\varepsilon}\right)\left(\frac{1}{\bar{p}_2 \cdot k-i\varepsilon}-\frac{(-1)^{n_2}}{\bar{p}_2 \cdot k+i\varepsilon}\right).
\end{multline}

We next make that for $x$ real,
\begin{equation}
{1\over x\pm i\varepsilon}= p.v.(x)\mp i \pi \delta(x)
\end{equation}
where $\varepsilon>0$ and $p.v.(x)$ is the principal value.
This implies that for $x$ real we have
\begin{equation}
    {1\over x+ i\varepsilon}+{1\over x- i\varepsilon}= 2p.v.(x);\qquad   {1\over x+ i\varepsilon}+{1\over x- i\varepsilon}= -2i \pi \delta(x).
\end{equation}
We now remark that for both $x$  and $y$ real,
\begin{multline}
  \left(\frac{1}{x+i\varepsilon}+\frac{1}{x-i\varepsilon}\right)
  \left(\frac{1}{y+i\varepsilon}+\frac{1}{y-i\varepsilon}\right)
=\left(\frac{1}{x+i\varepsilon}-\frac{1}{x-i\varepsilon}\right)\left(\frac{1}{y-i\varepsilon}-\frac{1}{y+i\varepsilon}\right)\cr
+\frac{2}{(x+i\varepsilon)(y+i\varepsilon)}+\frac{2}{(x-i\varepsilon)(y-i\varepsilon)}.
\end{multline}
Applied inside the box integral the last terms on the right-hand-side lead
to an imaginary contribution.
We therefore conclude that the real part of the box contribution is
given by
 \begin{equation}\label{e:reIone}
\Re( I_{\Box})=-\frac{|\vec{\underline q}|^{D-6}}{2
  \hbar^2}\left(\sum_{n=0}^{\infty} \sum_{m=0}^{2n} (-1)^m
\left(\frac{\hbar|\vec{\underline q}|}{4}\right)^{2n} \mathcal
B_{m,2n-m} \right)
\times \int \frac{d^D k}{(2\pi)^{D-2}}\frac{\delta(\bar p_1 \cdot
  k)\delta(\bar p_2 \cdot k)}{k^2 (k+u_q)^2}.
\end{equation}
Which is the soft expansion of the box integral in terms of the velocity cuts of~\cite{Bjerrum-Bohr:2021din}.

We remark that this expression indeed is the soft small $|\underline q|$
expansion of  the one-loop unitarity one-cut of the massive propagators
\begin{equation}\label{e:boxcut}
\Re( I_{\Box})=-{1\over2} I_{\Box}^{\rm 1-cut}\equiv-\frac{|\vec{\underline q}|^{D-6}}{2 \hbar^2}\int \frac{d^D
  k}{(2\pi)^{D-2}}\frac{\delta(2p_1\cdot k+k^2)\delta(-2p_2\cdot k+k^2)}{k^2 (k+u_q)^2}.
\end{equation}
Using {\tt LiteRed}~\cite{Lee:2013mka} to high orders in
$|\vec{\underline q}|^2$ we conjecture that
\begin{equation}
\sum_{m=0}^{2n} (-1)^m \mathcal B_{m,2n-m}=\frac{2^n s^n
  \prod_{j=0}^{n}(D-3-2j)}{(D-3)\big((\bar{p}_1\cdot \bar{p}_2)^2-\bar{p}_1^2\bar{p}_2^2\big)^n}.
\end{equation}
We have checked this expression by evaluating the expression up to and including $n=12$.

Performing the sum,
 \begin{equation}
\sum_{n=0}^{\infty} \prod_{j=0}^{n}\left(\frac{D-3}{2}-j\right)\Bigg(\frac{\hbar^2 |\vec{\underline q}|^2 s}{4\big((\bar{p}_1\cdot\bar{p}_2)^2-\bar{p}_1^2\bar{p}_2^2\big)}\Bigg)^{n} ={D-3\over2}\left(1+\frac{\hbar^2 |\vec{\underline q}|^2 s}{4((\bar{p}_1.\bar{p}_2)^2-\bar{m}_1^2\bar{m}_2^2)}\right)^{\frac{D-5}{2}},
\end{equation}
and we thus finally obtain the full real part of the box integral in the soft expansion: 
\begin{equation}
 I_{\Box}^{\rm 1-cut}=\frac{|\vec{\underline q}|^{D-6}}{4\hbar^2} \Big(1+\frac{\hbar^2 |\vec{\underline q}|^2 s}{4((\bar{p}_1.\bar{p}_2)^2-\bar{m}_1^2\bar{m}_2^2)}\Big)^{\frac{D-5}{2}}\int \frac{d^D k}{(2\pi)^{D-2}}\frac{\delta(\bar{p}_1 \cdot k)\delta(\bar{p}_2 \cdot k)}{k^2 (k+u_q)^2}.
\end{equation}
Noting that $(\bar{p}_1\cdot\bar{p}_2)^2-\bar{m}_1^2\bar{m}_2^2=m_1^2
m_2^2(\sigma^2-1- \frac{\hbar^2 |\vec{\underline q}|^2
  s}{4m_1^2m_2^2})$
and finally evaluating  the remaining bubble
integral we have thus established that the cut part of the box integral is given by
\begin{align}\label{e:boxrealfinal}
 I_{\Box}^{\rm 1-cut}&=\frac{|\vec{\underline q}|^{D-6}}{ \hbar^2}\int \frac{d^D
  k}{(2\pi)^{D-2}}\frac{\delta(2p_1\cdot k+k^2)\delta(-2p_2\cdot
              k+k^2)}{k^2 (k+u_q)^2},\\
\nonumber  &=\frac{|\vec{\underline q}|^{D-6}}{4 \hbar^2 m_1 m_2 \sqrt{\sigma^2-1}} \left(1- \frac{\hbar^2 |\vec{\underline q}|^2 s}{4m_1^2m_2^2(\sigma^2-1)}\right)^{\frac{4-D}{2}}\frac{\Gamma(\frac{D-4}{2})^2 \Gamma(\frac{6-D}{2})}{(4\pi)^{\frac{D-2}{2}} \Gamma(D-4)}.
\end{align}

\subsection{The two-cut part of the double-box integral}
Having gone through the derivation of the one-cut part of the one-loop box integral in such great detail we can be brief regarding the corresponding 
two-cut (or four velocity cut) computation of the doubly iterated tree in momentum space. We define
\begin{multline}
I_{\Box\Box}^{\rm 2-cut} ={1\over \hbar^{2D-7} }\int \frac{d^D l_1 d^D
  l_2}{(2\pi)^{2D-4}}
\frac{\delta((p_1-l_1)^2-m_1^2)\delta((p_2+l_1)^2-m_2^2)}{l_1^2
  (l_1+l_2-q)^2}\cr
\times\frac{\delta((p_1+l_2-q)^2-m_1^2)\delta((p_2-l_2+q)^2-m_2^2)}{ l_2^2}.
\end{multline}
Using $\bar{p}$-coordinates and neglecting the $l_i^2$ terms (for the same reason given above)
\begin{multline}
I_{\Box\Box}^{\rm 2-cut}={1\over\hbar^{2D-7}}\int \frac{d^D l_1 d^D
  l_2}{(2\pi)^{2D-4}} \frac{\delta(-2\bar{p}_1 \cdot l_1-q \cdot
  l_1)\delta(2\bar{p}_2 \cdot l_1-q \cdot l_1)}{l_1^2
   (l_1+l_2-q)^2}\cr
\times \frac{\delta(2\bar{p}_1 \cdot
  l_2-q \cdot l_2)\delta(-2\bar{p}_2 \cdot l_2-q \cdot l_2)}{
  l_2^2}.
\end{multline}
The soft series expansion leads to

\begin{multline}
I_{\Box\Box}^{\rm 2-cut}=\frac{1}{(2i\pi)^4\hbar^{2D-7}}\sum_{n_1=0}^{\infty}\sum_{n_2=0}^{\infty}\sum_{n_3=0}^{\infty}\sum_{n_4=0}^{\infty}\int
\frac{d^D l_1 d^D l_2}{(2\pi)^{2D-4}} \frac{(q \cdot l_1)^{n_1+n_2} (q
  \cdot l_2)^{n_3+n_4}}{l_1^2 l_2^2 (l_1+l_2-q)^2}\cr
\times\Big(\frac{1}{(-2\bar{p}_1 \cdot
  l_1-i\varepsilon)^{n_1+1}}-\frac{1}{(-2\bar{p}_1 \cdot
  l_1+i\varepsilon)^{n_1+1}} \Big) \Big(\frac{1}{(2\bar{p}_2 \cdot
  l_1-i\varepsilon)^{n_2+1}}-\frac{1}{(2\bar{p}_2 \cdot
  l_1+i\varepsilon)^{n_2+1}} \Big) \cr
\times\Big(\frac{1}{(2\bar{p}_1 \cdot l_2-i\varepsilon)^{n_3+1}}-\frac{1}{(2\bar{p}_1 \cdot l_2+i\varepsilon)^{n_3+1}} \Big)   \Big(\frac{1}{(-2\bar{p}_2 \cdot l_2-i\varepsilon)^{n_4+1}}-\frac{1}{(-2\bar{p}_2 \cdot l_2+i\varepsilon)^{n_4+1}} \Big).
\end{multline}
Computing the two first orders in $|\vec{q}|$ with {\tt LiteRed}~\cite{Lee:2013mka} gives
\begin{multline}
I_{\Box\Box}^{\rm 2-cut} = \left(1-\frac{(3+4\varepsilon)|\vec{ q}|^2 s}{12
  m_1^2 m_2^2 (\sigma^2-1)}\right){1\over16\hbar^{2D-7}}\int \frac{d^D
  l_1 d^D l_2}{(2\pi)^{2D-4}} \frac{\delta(\bar{p}_1 \cdot
  l_1)\delta(\bar{p}_1 \cdot l_2)\delta(\bar{p}_2 \cdot
  l_1)\delta(\bar{p}_2 \cdot l_2)}{l_1^2 l_2^2 (l_1+l_2-q)^2}\cr+\mathcal
  O(|\vec{q}|^{1-4\epsilon})
\end{multline}
which is evaluated to
\begin{multline}
I_{\Box\Box}^{\rm 2-cut} =
-\left(1-\frac{(3+4\varepsilon)\hbar^2|\vec{\underline q}|^2 s}{12
    m_1^2 m_2^2 (\sigma^2-1)}\right){1\over 16\hbar^3 (m_1^2
  m_2^2(\sigma^2-1)-\frac{\hbar^2|\vec{\underline q}|^2 s}{4})|\vec{\underline
    q}|^{2+4\varepsilon}}\cr
\times\frac{\Gamma(-\varepsilon)^3 \Gamma(1+2\varepsilon)}{(4 \pi)^{2-2\varepsilon} \Gamma(-3\varepsilon)}+\mathcal O(|\vec{q}|^{1-4\varepsilon}).
\end{multline}
Finally expanding also the $q^2$ of the denominator this gives the result quoted in the main text.

%%%%%%%%%%%%%%%%%%%%%%%%%%%%%%%%%%%%%%%

\end{document}